\documentclass{elsart}
\usepackage{epsfig}

\begin{document}
\begin{frontmatter}

\title{Deterministic Small-World Networks}

\author{Francesc Comellas}
\address{
Departament de Matem\`atica Aplicada IV,\\
Universitat Polit\`ecnica de Catalunya,\\
08034 Barcelona, Catalonia, Spain}
\author{Michael Sampels\corauthref{co}}
\corauth[co]{Corresponding author. Tel. +32-2-650-2711, Fax
+32-2-650-2715.}
\ead{msampels@ulb.ac.be}
\address{
Institut de Recherches Interdisciplinaires et de\\ 
D\'eveloppements en Intelligence Artificielle,\\
Universit\'e Libre de Bruxelles, CP 194/6,\\
Av. Franklin D. Roosevelt 50,\\
1050 Bruxelles, Belgium}

\begin{abstract}
Many real life networks, such as the World Wide Web, transportation
systems, biological or social networks, achieve both a strong local
clustering (nodes have many mutual neighbors) and a small diameter
(maximum distance between any two nodes). These networks have been
characterized as small-world networks and modeled by the addition of
randomness to regular structures. We show that small-world networks can
be constructed in a deterministic way. This exact approach permits a
direct calculation of relevant network parameters allowing their
immediate contrast with real-world networks and avoiding complex
computer simulations. 
\end{abstract}

\date{5 November 2001}
\end{frontmatter}


In 1967 Milgram \cite{Mil67} discovered in a now classical experimental
study that the median of the distances (counting the number of people on
acquaintance chains) between the inhabitants of the United States is
only 6. This fact was surprising, because it was contraintutitive to an
order of population of 200 million. However, from the theory of random
graphs \cite{BV82} it is known that distances in a randomly constructed
network with the same number of nodes are relatively small. Still, a
social network is not believed to be a totally randomly network, because
usually people knowing each other have a lot of other common
acquaintances. 

This led Watts \cite{W99} to define small-world networks as graphs which
have clustering coefficients much larger than random networks and with
diameters which increase logarithmically with the number of nodes. Watts
and Strongatz \cite{WS98} described stochastic models with which
small-world networks can be synthetically generated. Their central idea
is to begin with a very structured network such as a ring or a grid,
which has a high clustering but also a high distance between the nodes,
and to randomly add edges to this network in order to reduce the
diameter. The authors show that this addition of random noise to a
structured network leads first to a reduction of the distances in the
network and then to a reduction of the clustering. In the region
between, the resulting networks fulfill the above conditions of
small-world networks.

The introduction of this model for small-world networks has renewed the
interest on these networks. Many publications dealing with their
intrinsic characteristics such as connectivity distribution, robustness,
etc. have appeared recently~\cite{BA99,CNSW00,AJB00}. Small-world
networks have also been considered to model, for example, biological,
metabolic, transportation or social networks. A common point in most of
these studies is the use of randomness to construct a small-world
network. The degree distribution is Gaussian in the model of Watts and
Strongatz~\cite{WS98} and follows a power law (is scale-free) in the
model of Barab\'asi and Albert~\cite{BA99}. Recently, it has been
discovered that a constant degree distribution is also
possible~\cite{COP00}.

The introduction of deterministic models for networks intends to help in
the understanding of their behavior and permits the construction of
specific networks. A recent study~\cite{ASBS00} shows for example that
most real-world networks follow a power law with some deviations because
of their growing dynamics, and with deterministic models similar
networks could easily be synthesized. The question whether a scale-free
network can also be constructed in a deterministic manner has been
solved by Barab\'asi et al.~\cite{BRV01} by a hierarchical construction
scheme.

Here, we present two simple deterministic construction techniques for
small-world networks. These differ from real and randomly generated
small-world networks in that for a given diameter a small-world network
can explicitly be constructed, and different degree distributions are
possible, including networks with constant degree. This exact approach
permits a direct calculation of relevant network parameters and would
allow the design of specific small-world networks for applications such
as new communication systems and computer architectures.

\section{Deterministic models for small-world networks}

Our models are based on the replacement of nodes with networks and the
addition of new networks to an arbitrary low diameter ``backbone''
network resulting in small-world networks which have more nodes than the
original.

\subsection*{Constant degree small-world networks}

We consider first the replacement of nodes of a network by new networks.
Consider an arbitrary sparse network with $n$ nodes such that each node
has $k$ neighbors. Its expected diameter is $d \approx \log_{k-1}
n$~\cite{BV82} and the expected clustering coefficient is
$(k-1)/n$~\cite{W99}. This coefficient is defined as the fraction of
existing edges between the neighbors of a node and the maximum number of
edges that could possibly exist among the neighbors, averaged over all
the nodes of the network. A high clustering coefficient means that the
neighborhood of an average node is highly connected to each other.

For $n \gg k$ the clustering is almost zero. Replacing each node with a
totally connected network of $k$ nodes will lead to a new network with
$kn$ nodes, diameter $2d+1$ and clustering $(n-2)/n$ such that each node
also has $k$ neighbors (Fig.~1b). This new network has the
characteristics of a small-world network: a relatively low diameter and
a high clustering. Moreover, this construction maintains symmetry
properties of the original network. For example, if the original network
looks the same from any node and any link (is node-symmetric and
edge-symmetric) the resulting network is also node-symmetric. This
property is important, for instance in communication networks, because
it simplifies routing algorithms and network management. Barab\'asi et
al.~\cite{BRV01} already mention partial self-similarity for their
construction of scale-free networks.

\subsection*{Small-world networks with variable degree distributions}

In a second construction a small-world network is obtained by connecting
nodes of a network of diameter $d$ to a complete network of any size,
which may be different from node to node (Fig.~1c). In that case the
resulting network has diameter $d+2$, a high clustering and nodes may
have a different number of neighbors. The construction can be
generalized if we connect dense networks with reduced diameter ($\le
d'$) instead of complete networks. In that case we obtain a small-world
network with diameter at most $d+2d'$. This technique is very flexible
as it allows different final degree distributions, including scale-free
networks. Thus, it can be used for the construction of small-world
networks adapted for different contexts.

\section{Discussion}

With our models we show that it is quite simple to generate a network
with low diameter and high clustering. The World Wide Web, for example,
has been increasing in size exponentially from a few thousand nodes in
the early nineties to hundreds of millions today. However, it has been
shown recently ~\cite{AJB99} that the diameter of the Web is small and
grows very slowly -- a tenfold increase on the present size of the World
Wide Web would only modify its diameter from 19 to 21. Similar to
Milgram's study, this result might be astonishing at first sight. But
with our second model, starting from a network with 300,000 nodes,
diameter 15 and clustering 0 and such that each node has 5 neighbors, we
obtain, for example, a network with 3,300,000 nodes, diameter 17 and
clustering 0.95, by adding complete networks of 10 nodes to each node.
This is in contradiction to the usual belief that the World Wide Web has
an extremely good topology. It is easily possible to construct networks
with a higher clustering index and a lower diameter by our deterministic
techniques. Neither complex simulations nor stochastic models are
necessary to generate networks with similar small-world
characteristics.\footnote{Although the replacement of vertices by
networks in our model could be associated to a local growth of an
existing network, we do not intend to model the growth of real networks,
which is considered to be much more difficult. We only want to give an
immediately comprehensible technique how to generate a network with
small-world characteristics.}

It is possible to obtain a scale-free network with the same number of
nodes, diameter and clustering by adding complete networks of different
sizes and distributed according to a power-law. In the same manner, our
constructions can model the existence of small-world networks which have
similar defining parameters (size, clustering, diameter) but different
connectivity distributions such as: Gaussian, in the randomized model of
Watts and Strogatz \cite{WS98}; power-law distributions, in some real
life networks \cite{BA99} such as the power grid of the western US, the
collaboration graph of actors, etc.; and constant distributions, in
which all nodes of the small-world network have the same number of
neighbors \cite{COP00}. The addition of totally connected networks of
different size to the appropriate nodes can produce the desired
distribution. 

We believe that the variety of deterministic small-world networks that
may be obtained with these simple constructions will help to demystify
the small-world phenomenon. We claim that using deterministic
techniques it will be possible to construct networks which are much
better than those found in nature. As no simulation is needed in our
models and relevant parameters can be directly calculated, the contrast
of real with synthesized networks can be done immediately.

\section*{Acknowledgements}
Research supported by 
\emph{Comisi\'on Interministerial de Ciencia y Tecnologia}, Spain 
(TIC97-0963).


\begin{figure}[htbp]
\epsfig{figure=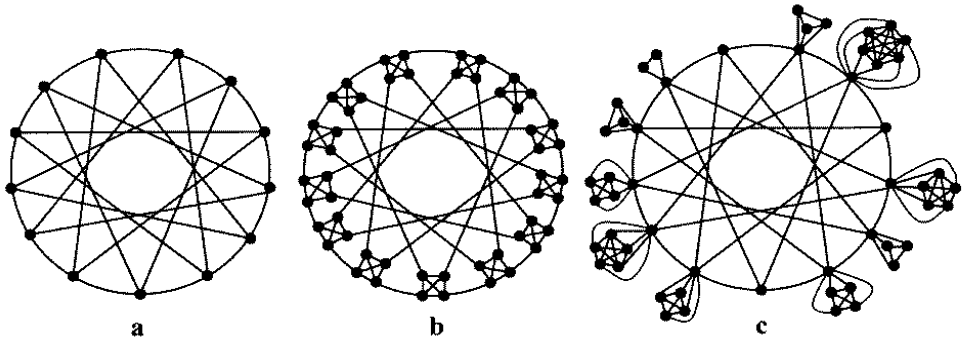,height=6cm,width=\textwidth}
\caption{{ (a)} The optimal node-symmetric network of 
diameter 2 with 4 neighbors per node has 13 nodes. 
{(b)} If each node is replaced by a complete 
network with 4 nodes, the resulting network is also 
node-symmetric, has 52 nodes and diameter 5. 
The clustering increases from 0 to 0.5. 
{ (c)} By connecting complete networks to each 
node of the original network we obtain a new network 
also with 52 nodes, but with diameter 4 and clustering 0.79. 
In this case nodes have different number of neighbors.}
\label{Fig1}
\end{figure}


\begin{thebibliography}{10}
\bibitem{Mil67} S. Milgram,
The Small-World Problem,
Psychology Today 1 (1967), 60--67.

\bibitem{BV82} B. Bollob\'as, and  F. de la Vega,
The diameter of random graphs, 
Combinatorica 2 (1982) 125--134.

\bibitem{W99} D.J. Watts,  
Small Worlds: The Dynamics of Networks between Order and Randomness,
Princeton University Press, Princeton, NJ, 1999.

\bibitem{WS98} D.J. Watts, and S.H. Strogatz, 
Collective dynamics of `small-world' networks,
Nature 393 (1998) 440--442. 

\bibitem{BA99} A.-L. Barab\'asi,  and R. Albert,
Emergence of scaling in random networks, 
Science 286 (1999) 509--512.

\bibitem{CNSW00} D.S. Callaway, M.E.J. Newman, S.H. Strogatz,  and D.J.
Watts, Network robustness and fragility: Percolation on random graphs,
Phys. Rev. Lett. 85 (2000) 5468--5471.

\bibitem{AJB00} R. Albert, H. Jeong, and A.-L. Barab\'asi,
Attack and error tolerance in complex networks,
Nature 406 (2000) 378--382. 

\bibitem{ASBS00} L.A.N. Amaral, A. Scala, M. Barth\'el\'emy, and H.E.
Stanley, Classes of small-world networks,
Proc. Natl. Acad. Sci. USA  97 (2000) 11149–-11152.

\bibitem{AJB99} R. Albert, H. Jeong, and A.-L. Barab\'asi,
Diameter of the World-Wide Web, 
Nature 401  (1999) 130--131. 

\bibitem{COP00} F. Comellas, J. Oz\'on, and  J.G. Peters,
Deterministic small-world\\ communication networks,
Inform. Process. Lett.  76 (2000) 83--90.

\bibitem{BRV01} A.-L. Barab\'asi, E. Ravasz, and T. Vicsek,
Deterministic scale-free networks, 
Physica A 299 [3-4] (2001)  559--564. 
\end{thebibliography}
\end{document}